\documentstyle[12pt]{article}

\begin{document}
\bigskip

\title{Relativistic Description of $J/\psi$  Dissociation\\ in Hot Matter(cq6346)}

\bigskip
\author{{
     Hong-shi Zong$^{1,2}$, Jian-zhong Gu$^{1,2,5}$, Xiao-fei Zhang$^{1,3}$,}\\
    {Yu-xin Liu$^{1,2,4}$ and En-guang Zhao$^{1,2}$ } \\
\normalsize{$^{1}$ CCAST ( World Lab.), P. O. Box 8730, Beijing 100080, P. R. China} \\
\normalsize{$^{2}$ Institute of Theoretical Physics, Academia Sinica,
P. O. Box 2735}\\
\normalsize{Beijing, 100080, P. R. China}\thanks{Mailing Address}\\
\normalsize{$^{3}$ Institute of High Energy Physics, Academia Sinica,
Beijing, 100039, P. R. China}\\
\normalsize{$^{4}$ Department of Physics, Peking University, Beijing 100871, P. R. China}\\
\normalsize{$^{5}$ Max-Planck-Institut fuer Kernphysik, Postfach 103980, D-69029 Heidelberg, Germany}}

\date{}
\maketitle

\abstract{The mass spectra and binding radii of heavy quark bound states
are studied on the basis of the reduced Bethe-Salpeter equation. The
critical values of screening masses for $c\bar{c}$ and $b\bar{b}$
bound states at a finite temperature are obtained and compared with
the previous results given by non-relativistic models.}

\bigskip

\bigskip

PACS number(s): 13.85.Ni, 25.75.Dw, 12.38.Mh

\bigskip

\newpage

\begin{center}
\section{Introduction}
\end{center}
   One of the main aims of high energy nuclear collisions is to
explore a new state of matter, the Quark-Gluon Plasma(QGP),
through heavy ion collisions in the laboratory.
It was theoretically proposed  that a suppression of $J/\psi$ production in
relativistic heavy ion collisions can serve as a clear signature
for the formation of QGP [1].
Subsequently
this suppression effect was observed by NA38 collaboration [2].
However, successive research pointed out
that such suppression could also
exist in hadronic matter (HM), even though caused by a completely different
mechanism [3]. The anomalous $J/\psi$ suppression
has recently been reported by the NA50 collaboration [4]
and there have been
a number of attempts to explain it [5-7].
Some authors believe that the data may implicate the possibility of the
formation of a QGP [8]. For understanding the experimental data clearly, the
dissociation mechanism of $J/\psi$ in hot QGP must also be
studied carefully.

In QGP, quarks and gluons are deconfined and the confining force
between quark and antiquark vanishes, the only interaction between
quark and antiquark is the Coulomb-type color interaction. The color
charge of a quark will be screened by the quark sea in the plasma.
Due to Debye screening the final yields of $J/\psi$ will be
suppressed. Binding and dissociation of $J/\psi$ at a finite temperature have
been studied in the non-relativistic formalisms [9,10].
The $J/\psi$ was regarded
as a non-relativistic bound state in those papers.
However, generally speaking,
the motion of a quark and an antiquark in a meson is relativistic, even for
charmonium. As pointed out in Ref.[11],
for the $c \bar{ c}$ system, the kinetic energy
is about $13\%$ of the total energy and the ratio of
relativistic corrections
to the quark mass will not decrease with the increase of quark mass if the
interaction is of Coulomb-type. As a result
the bound state equation for $J/\psi$
in general should be relativistic. So it is an interesting task to discuss
the binding and dissociation of
charmonium in hot matter in a relativistic
formalism. This is the main purpose of this paper.

It is well known that
the Bethe-Salpeter (BS) equation [12] is
the only effective
relativistic equation of the two-body bound state problems.
Because of its consistency with
quantum field theory, the BS equation can be used for the study of the
binding and
dissociation of charmonium and bottonium
which has seldom investigated in the framework of
relativistic formalism in a thermal environment and of great interest.
We expect that our calculations of observables could readily yield
results at variance with those of non-relativistic models. Furthermore,
we are not only interested in the
energy spectrum of mesons, which is an important source to study
the interquark dynamics, but also in the wave functions which play a
key role in the calculations of the root-mean-square(r.m.s) radii
of $c\bar{c}$ and $b\bar{b}$
bound states.

In this paper, we shall discuss the binding and
dissociation of heavy quark resonances in the hot matter within the context
of the BS equation. In section 2, we focus our attention on the interaction
between quark
and antiquark in mesons and the properties of the BS equation. In section 3,
we use the BS equation to calculate the mass spectra, r.m.s. radii and
critical values of the screening masses for the
$c\bar{c}$ and $b\bar{b}$ bound states  
and compare them with the previous results. The sensitivity
of our results to the Lorentzian structure of the confining
potential is also checked. In section 4, we discuss the results and conclude. 

\begin{center}
\section{Formalism}
\end{center}

It is well known that the BS equation is a proper tool for
describing the relativistic two-body
bound state problems[13]. The full bound state BS equation in
momentum space, written in the two-sided notation, reads
\begin{equation}
(\eta_{1}\rlap/P+\rlap/p-m_{1})\chi_{P}(p)(\eta_{2}\rlap/P-\rlap/p+m_{2})=
\frac{i}{(2\pi)^{4}}\int~d^{4}p'V(p,p';P)\chi_{P}(p')
\end{equation}
where $\eta_{i}=\frac{m_{i}}{m_{1}+m_{2}}$ (i=1,2), $\chi_{P}$
is the momentum-space wave function for the quark-antiquark
system with total four momentum P in momentum space, p is the
relative four momentum. V is the interaction kernel that acts
on $\chi_{P}$ and formal products of V$\chi_{P}(p')$
in Eq.(1) take
the form V$\chi_{P}(p')=V_{s}\chi_{P}(p')
+\gamma_{\mu}\otimes\gamma^
{\mu}V_{v}\chi_{P}(p')$, in which $V_{s}$ and $V_{v}$ are scalar, vector
potential respectively.
The short-distance behaviour of the V can be calculated in QCD using
perturbative theory. However, the long-distance behaviour of the V involves
non-perturbative effects, and the Lorentzian structure of the confining potential
is not established theoretically in QCD. Consequently, we shall treat the
form of V in a partially
phenomenological way. 
The parameters $m_{1}$, $m_{2}$ should be interpreted as effective constituent masses and
similarly the whole propagator is an effective one.

Using the standard reduction and spin-independent treatment
, one can get the spin-independent reduced Salpeter equation [14]
for the three-dimensional equal-time BS wave function
\begin{equation}
\phi(\vec{p})=\int~dp^{0}\chi_{P}(p^{0},\vec{p}),
\end{equation}
\begin{equation}
(M-E_{1}-E_{2})\phi(\vec{p})=\int~\frac{d^{3}p'}{(2\pi)^{3}}
\sum_{i=s,v}~F_{i}^{si}(\vec{p},\vec{p'})V_{i}(|\vec{p}-\vec{p'}|)
\phi(\vec{p'}),
\end{equation}
where M is the mass of a $q\bar{q}$ bound state, $E_{i}=
(\vec{p}^{2}+m_{i}^{2})^{\frac{1}{2}}$, i=1,2 represent a quark
and an antiquark, respectively.  The functions $F_{v}^{si}$
and $F_{s}^{si}$ appearing in Eq.(3) are

\begin{eqnarray}
F_{v}^{si}(\vec{p},\vec{p'})& =& \frac{1}{4E_{1}E_{2}}[(E_{1}+m_{1})
(E_{2}+m_{2})+\vec{p}^{2} +
\frac{(E_{1}+m_{1})(E_{2}+m_{2})}{(E_{1}'+m_{1})(E_{2}'+m_{2})}
\vec{p'}^{2} \nonumber \\ \noalign{\vskip2mm}
& &  +\!\frac{(\vec{p}\cdot\vec{p'})^{2}}
{(E_{1}'\!+\!m_{1})(E_{2}'\!+\!m_{2})}+\!
(\frac{E_{1}\!+\!m_{1}}{E_{2}'\!+\!m_{2}}\!+\!\frac{E_{2}\!+\!m_{2}}
{E_{1}'\!+\!m_{1}}\!+\!\frac{E_{1}\!+\!m_{1}}{E_{1}'\!+\!m_{1}}\!+\!
\frac{E_{2}\!+\!m_{2}}{E_{2}'\!+\!m_{2}})\vec{p}\cdot\vec{p'} ], \nonumber \\
& & 
\end{eqnarray}

and
\begin{equation}
F_{s}^{si}(\vec{p},\vec{p'})=\frac{1}{4E_{1}E_{2}}[
(E_{1}+m_{1})(E_{2}+m_{2})-(\frac{E_{1}+m_{1}}{
E_{2}'+m_{2}}+\frac{E_{2}+m_{2}}{E_{1}'+m_{1}})
\vec{p}\cdot\vec{p'}+\frac{(\vec{p}\cdot\vec{p'})^{2}}{
(E_{1}'+m_{1})(E_{2}'+m_{2})}].
\end{equation}
Here $E_{i}'=(\vec{p'}^{2}+m_{i}^{2})^{\frac{1}{2}}$, i=1,2.
Since $F_{v}^{si}$ and $F_{s}^{si}$ are
spin-independent, the singlet and triplet are degenerate. There
is no coupling between different orbital angular momenta in Eq.(3).
We can therefore extract the angular dependence of $\phi(\vec{p})$
in a single spherical harmonic basis
\begin{equation}
\phi(\vec{p})=\phi_{nL}(|\vec{p}|)Y_{Lm}(\stackrel{\wedge}{\vec{p}}).
\end{equation}
By using the following identity
\begin{equation}
\frac{4\pi}{2L+1}\sum_{m}Y_{Lm}^{*}(\stackrel{\wedge}{\vec{p}})
Y_{Lm}(\stackrel{\wedge}{\vec{p'}})=P_{L}(cos\theta),
\end{equation}
where $\theta$ is the angle between the $\stackrel{\wedge}{
\vec{p}}$ and $\stackrel{\wedge}{\vec{p'}}$,
one can obtain the following equation for the radial wave function
$\phi_{nL}(|\vec{p}|)$
\begin{equation}
(M_{nL}-E_{1}-E_{2})\phi_{nL}(|\vec{p}|)=
\int~\frac{d^{3}\vec{p'}}{(2\pi)^{3}}\sum_{i=s,v}F_{i}^{si}
(\vec{p},\vec{p'})V_{i}(|\vec{p}-\vec{p'}|)P_{L}(cos\theta)
\phi_{nL}(|\vec{p'}|).
\end{equation}
Eq.(8) gives a well-defined eigenvalue problem for the masses
of the $q\bar{q}$ bound states in momentum space. Here we would like
to point out that the momentum dependence of the interaction is
treated exactly in the above equation.

 In the non-relativistic limit, Eq.(8) can be reduced to 
the usual Schr\"{o}dinger equation
\begin{equation}
(M_{nL}-E_{1}-E_{2})\phi_{nL}(|\vec{p}|)=
\int~\frac{d^{3}\vec{p'}}{(2\pi)^{3}}
[V_{s}(|\vec{p}-\vec{p'}|)+V_{v}(|\vec{p}-\vec{p'})]P_{L}(cos\theta)
\phi_{nL}(|\vec{p'}|).
\end{equation}

It will be convenient in calculating the r.m.s radii of $q\bar{q}$
bound state to transform $\phi_{nL}(|\vec{p}|)$ to the position space.
Making use of the following identity
\begin{equation}
\exp(i\vec{p}\cdot\vec{r})=4\pi\sum_{l=0}^{\infty}i^{l}j_{l}(p r)
Y_{lm'}^{*}(\Theta,\Phi)Y_{lm'}(\theta,\varphi),
\end{equation}
where the direction of $\vec{p}$ and $\vec{r}$ is specified by the polar
angles ($\Theta,\Phi$) and ($\theta,\phi$), respectively, we get
\begin{equation}
\begin{array}{ccc}
\phi_{nLm}(\vec{r})=\frac{1}{(2\pi)^{3}}\int~exp(i\vec{p}\cdot\vec{r})
\phi_{nLm}(\vec{p})d^{3}\vec{p}\\
=\frac{i^{L}}{2\pi^{2}}\int~p^{2}dp~j_{L}(pr)\phi_{nL}(|\vec{p}|)
Y_{Lm}(\theta,\phi).
\end{array}
\end{equation}
where we have used the orthogonal relation of spherical harmonic function and
Eq.(7). According to Eq.(10) one can get the radial wave function
in the position space
\begin{equation}
\phi_{nL}(|\vec{r}|)=\frac{i^{L}}{2\pi^{2}}\int~p^{2}dp
j_{L}(pr)\phi_{nL}(|\vec{p}|)
\end{equation}
in which the special functions $j_{0}(x)=\frac{sinx}{x}$,
$j_{1}(x)=\frac{sinx}{x^{2}}-\frac{cosx}{x}$.
Solving Eq.(8) and Eq.(12), one can obtain the masses
and corresponding r.m.s radii of the bound states.

To solve Eq.(8), one must have a good knowledge of the potential
between quark and antiquark. At present, it is commonly accepted that the
interaction between quark and antiquark consistes of a short-range part
describing the one-gluon-exchange(OGE) potential and an infinitely rising
long-range part responsible for the confinement of the quarks. As it is
well known that the OGE potential is a pure vector interaction. However
the Lorentzian structure of the confining interaction is not clear. Wilson loop
technique suggests that the confining potential should be taken purely
scalar[15], but relativistic potential calculations [16-17]
showed a need for some vector confinement. Therefore, we choose a
confining potential to be a mixture of a scalar and a vector[18]. This
leads to the following potential
\begin{equation}
V(r)=V_{s}(r)+V_{v}(r), 
\end{equation}
with
\begin{displaymath}
V_{s}=(1-x)\sigma~r,
\end{displaymath}
\begin{displaymath}
V_{v}=x\sigma~r-\frac{4}{3}\frac{\alpha_{s}}{r}.
\end{displaymath}
where $\sigma$ is the string tension, $\alpha_{eff}=-\frac{4}{3}
\alpha_{s}$ the effective coupling constant and x the vector-scalar mixing
parameter obey the condition $0 \leq x \leq 1$. Note that the edge of the interval
$x=0$ corresponds to the case of pure scalar confinement.

In a thermodynamical environment of interacting light quarks and
gluons at temperature T, quark binding becomes modified by color
screening [1]
\begin{equation}
V(r,\mu)=\left[x\frac{\sigma}{\mu}(1-e^{-\mu~r})-
\frac{4}{3}\frac{\alpha_{s}}{r}e^{-\mu~r}\right]+
\frac{\sigma}{\mu}(1-x)(1-e^{-\mu~r}).
\end{equation}
Here $\mu$ is the Debye screening mass (which is assumed to be a
function of temperature T) and the Debye screening length $r_{D}$ is defined
as the inverse of the screening mass, $r_{D}=\frac{1}{\mu}$. It is
necessary to note that the factor $\exp(-\mu~r)$ not only reflects
the color screening effects but also avoids the infrared divergence.
In fact, as pointed out in Ref.[19], the color screening effects
are also required at T=0 in order to fit the experimental properties
of quark systems. In this paper, we use the color screening potential
(Eq.(14)) to study the binding and deconfinement of heavy quark resonances
. In momentum space the potential can be written as
\begin{equation}
V_{s}(|\vec{p}-\vec{p'}|)=(1-x)
\left[\frac{\sigma}{\mu}\delta^{3}(
\vec{p}-\vec{p'})-\frac{\sigma}{\pi^{2}}\frac{1}{
[(\vec{p}-\vec{p'})^{2}+\mu^{2}]^{2}}\right],
\end{equation}

and
\begin{equation}
V_{v}(|\vec{p}-\vec{p'}|)=-\frac{2}{3\pi^{2}}\frac{\alpha_{s}}
{[(\vec{p}-\vec{p'})^{2}+\mu^{2}]}+x
\left[\frac{\sigma}{\mu}\delta^{3}(
\vec{p}-\vec{p'})-\frac{\sigma}{\pi^{2}}\frac{1}{
[(\vec{p}-\vec{p'})^{2}+\mu^{2}]^{2}}\right].
\end{equation}
The constants $\sigma$, $\mu$, $x$ and $\alpha_{s}$ are the parameters 
characterizing the potential.

\begin{center}
\section{Calculations and Results}
\end{center}

Based on the formula above, we first calculate the mass spectra
and r.m.s radii of $c\bar{c}$ and $b\bar{b}$ bound states with the
vector-scalar mixing parameter $x=0$, which corresponds to the pure scalar
confinement. The numerical
results are listed in Table I. The data used in our studies
consisted of the spin-averaged masses of $b\bar{b}$ and $c\bar{c}$
states and are given in the third column in Table I, with
\begin{equation}
\overline{M}_{nl}=\frac{1}{4(2l+1)}\sum_{j}(2j+1)M(n,j,l,s).
\end{equation}
Here we restrict ourselves to the first two radial excitations,
corresponding to the (spin-averaged) $J/\psi$ and $\Upsilon$ for n=1, $l$=0,
to the $\psi'$ and $\Upsilon$ for n=2, $l$=0, and to the $\chi_{c}$ and
$\chi_{b}$ for n=2, $l$=1.
Therefore $\overline{M}_{nl}$ can be written explicitly as:
$\overline{M}_{n0}=\frac{1}{4}[3M(n^{3}S_{1})+M(n^{1}S_{0})]$,
$\overline{M}_{n1}=\frac{1}{12}[5M(n^{3}P_{2})+3M(n^{3}P_{1})+
3M(n^{1}P_{1})+M(n^{3}P_{0})]$ (n=0,1 corresponding to
the first two radial excitations), where we recall the usual
spectroscopic notation $^{2s+1}L_{J}$ for a state with orbital
angular momentum L, spin s, and total angular momentum J; S, P, ...
correspond to orbital angular momentum L=0,1,..., respectively.
Since the spin-singlets $b\bar{b}(^{1}S_{0})$ and $b\bar{b}(^{1}P_{1})$
have not yet been unambiguously confirmed by experiment, we have
therefore used the results of previous spin-dependent fits to the data
to estimate the centers of gravity of the incomplete multiplets [14].
However, Ref.[10] compared their numerical results with the spin-triplet
rather than the spin-averaged masses of $b\bar{b}$ and $c\bar{c}$ systems.

   Table I. The mass spectra and rms radii of the $c\bar{c}$ and $b\bar{b}$
bound states.

\begin{center}
\begin{tabular}{|c|c|c|c|c|c|c|c|c|}
\hline
\multicolumn{1}{|c|}
{}  & {nl} &{Data}&\multicolumn{3}{|c|}{$M_{nl}(GeV)$} &
\multicolumn{3}{|c|}{$<r^{2}>^{\frac{1}{2}}$(fm)} \\
\hline
{ } &{ } &{ } &{Ref.[9]}
&{Ref.[10]} &{Ours}
&{Ref.[9]}&{Ref.[10]} & {Ours} \\
\hline
$c\bar{c}$  & 1S
& $J/\psi$(3.068) &3.0697&3.0700& 3.067 &0.4490&0.4453&0.2868\\
\hline
  & 2S
&$\psi^{'}$(3.663)&3.6978&3.6863&3.663&0.8655&0.9034&0.6317 \\
\hline
 &
3S&$\psi^{''}$(4.025)&4.1696&4.0806&4.019&1.2025&1.3765&0.8290\\
\hline
&1P&
$\chi_{c}$(3.525)&3.5003&3.5054 &3.526&0.6890&0.7000&0.5144   \\
\hline
$b\bar{b}$ & 1S&$\Upsilon$(9.436)
&9.4450&9.4310&9.436&0.2249&0.2211&0.1873\\
\hline
&2S &$\Upsilon^{'}$(10.013)&10.0040&10.0083&10.013&0.5040&0.4998&0.4480\\
\hline
&3S&$\Upsilon^{''}$(10.341)&10.3547&10.3564&10.343&0.7336&0.7457&0.6749
\\
\hline
&1P&$\chi_{b}$(9.899)&9.8974&9.8981&9.901&0.4041&0.3982&0.3569\\
\hline
\end{tabular}
\end{center}

In our calculations, we have used the following parameters.
$\sigma$=0.22 GeV$^{2}$, $\mu_{0}$=0.06 GeV, $m_{c}$=1.474 GeV,
$m_{b}$=4.762 GeV, $\alpha_{s}(c\bar{c})$=0.47,
and $\alpha_{s}(b\bar{b})$=0.38.  All these parameters are
within the scope of customary usage.
According to the concept of a running gauge coupling constant,
in the computations we allow different values of $\alpha_{s}$
for charmonium ($\alpha_{s}(c\bar{c})$) and bottonium ($\alpha_{s}
(b\bar{b})$) [20]. The relative magnitude of $\alpha_{s}$ is in accordance
with the idea of asymptotic freedom as expected for the strong gauge
coupling constant of quantum chromodynamics, that is, $\alpha_{s}(
b\bar{b})$$<$$\alpha_{s}(c\bar{c})$.

In order to compare our results with those given in Refs.[9-10],
we calculate the quantity $\chi^{2}$ which is defined as
\begin{equation}
\chi^{2}=\frac{\sum_{nl}(M_{nl}^{exp}-M_{nl}^{theory})^{2}}{
N-1}
\end{equation}
with N being the total number of ($nl$) state. We obtain  $\chi$=
0.0066 GeV for our numerical results. Comparing with $\chi$=
0.0223 GeV (for Ref.[10]) and $\chi$=0.0511 (for the Ref.[9]),
one can observe that the mass spectra obtained at present are more
consistent with the experiment than the previous results.

As mentioned above, the wave functions play a key role in the
calculation of r.m.s radii of $c\bar{c}$ and $b\bar{b}$ systems
. The last column in Table I shows that the r.m.s radii of $c\bar{c}$
and $b\bar{b}$ bound states given by our calculations are smaller
than those of Refs.[9,10]. This means that $J/\psi$ is
more tightly bound in our case than estimated by non-relativistic
models.

Next, we study the dissociation of $b\bar{b}$ and $c\bar{c}$ systems.
A suitable quantity to observe the vanishing of bound states is
the dissociation energy
\begin{equation}
E^{nl}_{dis}(\mu)=m_{1}+m_{2}+\frac{\sigma}{\mu}-M_{nl}(\mu).
\end{equation}
The dissociation energy is positive for bound states and turns 
negative for the continuum. Thus
\begin{equation}
E^{nl}_{dis}(\mu_{c})=0
\end{equation}
defines the critical value of $\mu$ beyond which there is no bound state
for the given quantum numbers. The calculated results of
the $E^{nl}_{dis}(\mu_{c})$ for the $c\bar{c}$ and $b\bar{b}$ systems
are given in Fig.1 and Fig.2 respectively.
The figures show that the dissociation energies
of our calculations (solid lines) are shifted
to larger $\mu$ regions in comparision with those (dotted lines)
of Ref.[10]. The reason is that our calculations are based on
the relativistic formula in which the momentum dependence is treated exactly,
which is different from that used in Ref.[10].

The calculated critical values of Debye masses and $M_{nl}(\mu_{c})$
for $c\bar{c}$ and $b\bar{b}$ resonances 
are given in Table II.

Table II. The calculated $\mu_{c}$ and $M_{nl}^{c}$ for charmonium
($c\bar{c}$) and bottonium ($b\bar{b}$).

\begin{center}
\begin{tabular}{|c|c|c|c|c|c|c|c|}
\hline
\multicolumn{1}{|c|}
{}  & {state} &\multicolumn{3}{|c|}{$\mu_{c}$(GeV)}
&\multicolumn{3}{|c|}{$M_{nl}(\mu_{c})$(GeV)}
\\
\hline
{}  & {} &{Ref.[9] }&{Ref.[10] } & {Ours} &{Ref.[9] }
& {Ref.[10]} & {Ours}\\
\hline
Charmonium  & 1S
&0.700  &0.600  &0.900 &2.9145 &2.8779 &3.1911\\
\hline
  & 2S
&0.360 &0.260 &0.470 &3.1725 &3.2964 &3.4160\\
\hline
 &
1P&0.342 &0.242 &0.450 &3.1982 &3.3513 &3.4363\\
\hline
Bottonium&1S&
1.560 &1.500 &1.640 &9.6108 &9.5379 &9.6581  \\
\hline
  & 2S&
0.660&0.560 &0.690 &9.7838 &9.7528 &9.8426\\
\hline
&1P &0.578 &0.460 &0.640 &9.8226 &9.8274 &9.8670\\
\hline
\end{tabular}
\end{center}

 Table II shows that the critical values of the screening masses for
$c\bar{c}$ and $b\bar{b}$ dissociations given by our calculations
are larger than those given by Refs.[9,10]. This indicates that
the results are dependent on the model. So a finer calculation
of the screening masses for $c\bar{c}$ and $b\bar{b}$
dissociation is needed
. Because the $J/\psi$ suppression is related to the colour screening,
the dissociation of $J/\psi$ is more interesting at finite temperatures.
According to our calculations, the critical value of the screening mass
for $J/\psi$ dissociation is about $\mu_{c}$=0.900 GeV (the corresponding screening
length is 0.219 fm). This information is probably useful to the study and
observation of $J/\psi$ production in high energy collisions.

In view of Table II, we would like
to further note that the masses of all bound
states are affected slightly by the change of $\mu$. 
As shown by Refs.[9,10], with the increment of $\mu$ the masses of the $c\bar{c}$ bound states
and those of the higher $b\bar{b}$ bound states decrease, 
while the $\Upsilon$ mass increases, which is different from ours.
In our case, the masses of the higher $c\bar{c}$ and
$b\bar{b}$ bound states decrease with $\mu$, while the $J/\psi$
and $\Upsilon$ masses increase with $\mu$. This means that the positive string tension
part of the potential dominates and is reduced as $\mu$ increases; only for
the $\Upsilon$ and $J/\psi$
does the second term in the right hand side of Eq.(14) give the main contribution in the
relativistic formalism.

One can also find in Table II that the relativistic correction for
$c\bar{c}$ bound states is larger than that of $b\bar{b}$ ones.
This is not surprising. Even before the detailed numerical
calculations, this qualitative conclusion can be reached based on
the following reasonable physical consideration.

As mentioned above, the Schr\"{o}dinger equation is the
non-relativistic limit of the BS equation. Usually, the b quark is heavy enough
compared to $\Lambda_{QCD}$, and one can expect that the relativistic correction
is small for $b\bar{b}$ bound states. Nevertheless, the mass of c quark
, $m_{c}$, is not much larger than $\Lambda_{QCD}$. The relativistic
correction may be large in the case of $c\bar{c}$ bound states, which is corroborated by
the detailed numerical calculations listed in Table II.

Finally, we check the sensitivity of our results with respect to the
vector-scalar mixing parameter $x$ appearing in the potential, the numerical results
are listed in Table III.

On the basis of the analysis of the numerical results (see Table III), we come
to the following conclusions:

For the systems containing two heavy quarks the sensitivity of the mass
spectra to the Lorentz structure of the confining potential is rather
moderate, especially in the $b\bar{b}$ system. Therefore, our results based
on pure scalar confining potential are insensitive to the particular
Lorentz structure of confinement.

Table III. The dependence of $c\bar{c}$ and $b\bar{b}$ masses (GeV) on the
mixing parameter $x$.

\begin{center}
\begin{tabular}{|c|c|c|c|c|c|c|c|c|}
\hline
{}  & {states} &{$x=0.0$} &{$x=0.1$} &{$x=0.3$} &{$x=0.5$}
&{$x=0.7$} &{$x=0.9$} &{$x=1.0$}
\\
\hline
{$c\bar{c}$}  & {1S} &{3.067}&{3.085} &{3.120} &{3.155}
& {3.189} & {3.223} &{3.240}\\
\hline
{}  & {2S}
&3.663  &3.686  &3.730 &3.773 &3.814 &3.854 &3.874\\
\hline
{}  & {3S}
&4.019 &4.053 &4.118 &4.180 &4.238 &4.293 &4.320\\
\hline
 &
1P&3.526 &3.552 &3.602 &3.650 &3.698 &3.745 &3.768\\
\hline
{$b\bar{b}$}&1S&
9.436 &9.440 &9.448 &9.455 &9.462 &9.470 &9.473 \\
\hline
  & 2S&
10.013&10.017 &10.027 &10.036 &10.047 &10.053 &10.058\\
\hline
  & 3S&
10.343&10.350 &10.363 &10.376 &10.390 &10.402 &10.408\\
\hline
&1P &9.901 &9.905 &9.915 &9.924 &9.933 &9.943 &9.947\\
\hline
\end{tabular}
\end{center}

\begin{center}
\section{Discussions and Conclusions}
\end{center}
  In summary, we have studied the mass spectra, root-mean-square radii and
dissociation of $c\bar{c}$ and $b\bar{b}$ in the relativistic formalism.
We would like to point out that a relativistic treatment of quark
-antiquark bound states, by means of the reduced BS equation, does
imply some improvement in the description of $c\bar{c}$ and $b\bar{b}$
meson mass spectra. This indicates that the relativistic description
is necessary for a finer calculation of $J/\psi$ at finite temperatures.
Because the $m_{c}$ is not heavy enough compared to $\Lambda_{QCD}$,
the relativistic effect can not be neglected for the description of $J/\psi$
dissociation in hot matter.
The critical values of the screening masses have also been calculated
and compared with the previous results given by the non-relativistic models.
The critical value of the screening mass for $J/\psi$ dissociation is
$\mu_{c}$=0.600 GeV in Ref.[10] and $\mu_{c}$=0.700 GeV
in Ref.[9], respectively.
In the present calculations, however, $\mu_{c}$=0.900 GeV,
which is larger than those from non-relativistic
models. This means that the magnitude of the screening masses is
model dependent. In order to get a finer evaluation of the screening mass
of $J/\psi$, one must have a good knowledge the confining potential
in addition to a relativistic description.
Therefore the study of quark confining potential
is of importance and should be further proceeded.

 Stimulating discussions with Weiqing Zhao and Wanyun Zhao are gratefully
acknowledged by one of the authors (H.Zong). We thank Baltin Reinhard for carefully
reading our manuscript.


\newpage

\noindent{\large\bf Figure Captions}

\begin{description}
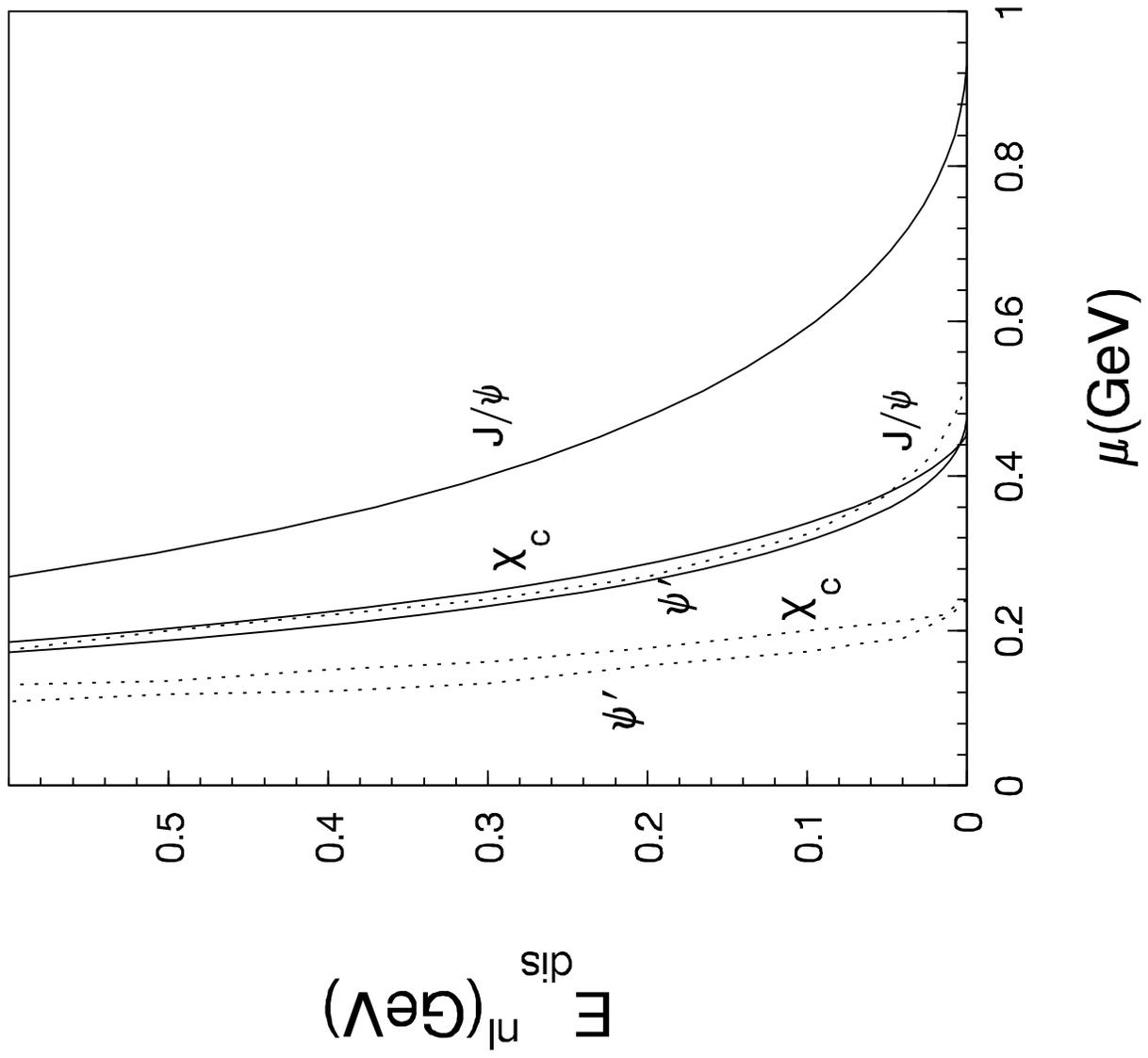
\item{Fig.1} The dissociation energies for $c\bar{c}$ bound states.
       The solid lines indicate our results, the dotted ones are from Ref.[10].
\vskip1cm
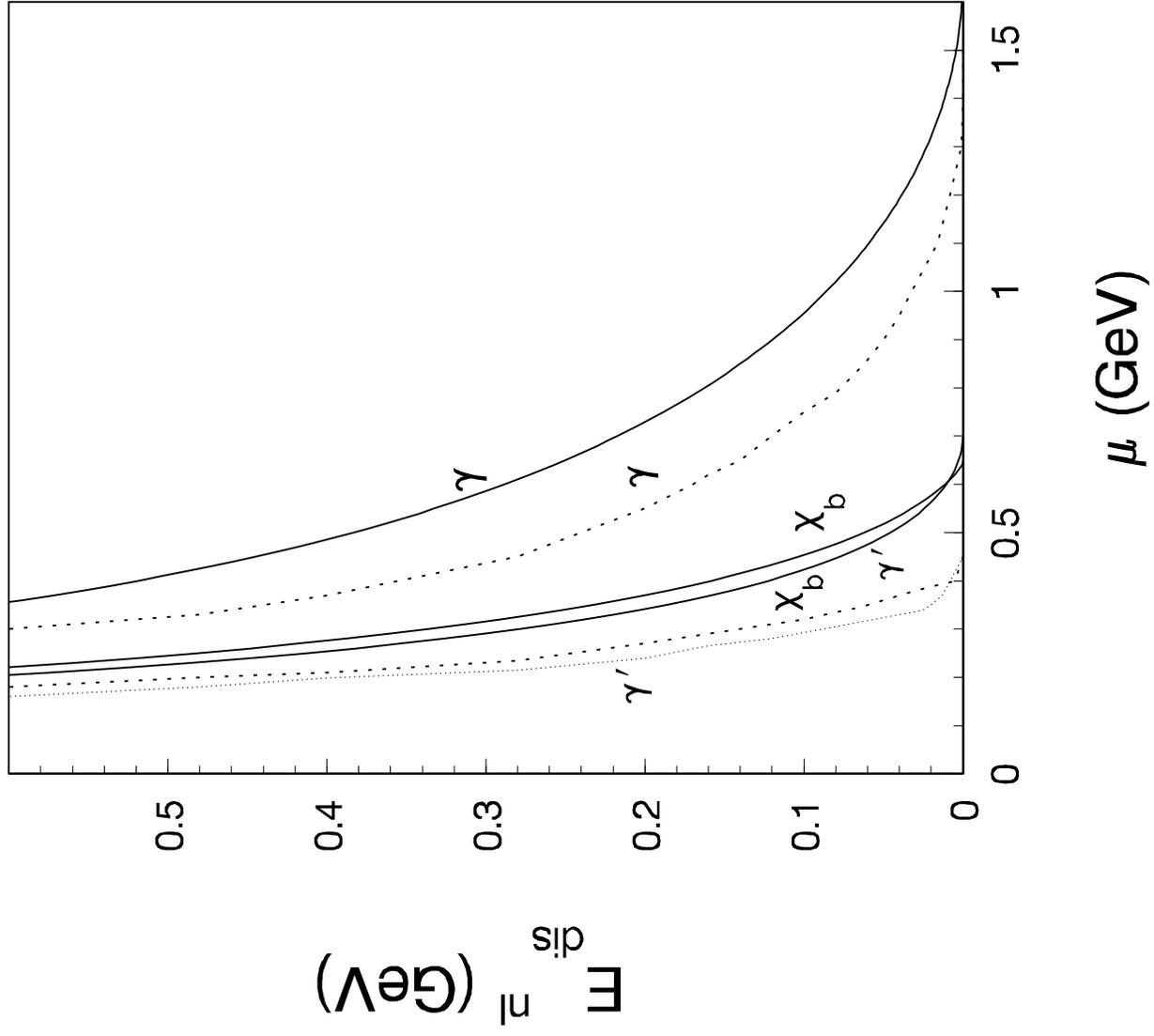
\item{Fig.2} The dissociation energies for $b\bar{b}$ bound states.
       The solid lines represent our results, the dotted ones are from Ref.[10].
\end{description}

\end{document}